\documentclass{aastex}
\slugcomment{The Astronomical Journal, in press}
\shorttitle{Shapley Supercluster: spectroscopic observations}
\shortauthors{Quintana et al.}

\def\deg{${}^\circ$\phm{ }}

\def\hh{${}^{h}$\phm{ }}
\def\mm{${}^{m}$\phm{ }}

\def\MM{${}^{m}$\llap{.}}

\begin{document}

\title{The Shapley Supercluster. II. Spectroscopic observations in a wide area 
and general morphology}

\author{H. Quintana\altaffilmark{1}, Eleazar R. Carrasco\altaffilmark{2}, and 
Andreas Reisenegger\altaffilmark{1}}

\altaffiltext{1}{Departamento de Astronom\'{\i}a y Astrof\'{\i}sica, Pontificia 
Universidad Cat\'olica de Chile, Casilla 306, Santiago 22, Chile.}
\altaffiltext{2}{Instituto Astron\^omico e Geof\'{\i}sico, Universidade de S\~ao 
Paulo, Caixa Postal 3386, 01060-970, S\~ao Paulo, Brazil.}

\begin{abstract}
We report 2868 new multi-object spectroscopic measurements of galaxy redshifts
in an area roughly 12\deg$\times$ 6\deg (right ascension $\times$ declination)
centered on the Shapley  Supercluster (SSC). These correspond to 2627 different
galaxies. Including  other measurements reported in the literature, the total
number of galaxies with measured redshifts in a 19\deg$\times$ 16\deg area
centered on the supercluster now reaches 5090. Of these, 2949 lie in the
redshift range 9,000 - 18,000 km s$^{-1}$, which we tentatively identify with
the SSC.  This unprecedented sample allows a quite detailed qualitative
morphological study of the SSC. Based on the three-dimensional distribution of
galaxies  and clusters of galaxies, we identify several sub-condensations of the
supercluster, as well as walls and filaments connecting them. We also find
another supercluster in the background, at redshift $\sim$ 23,000 km s$^{-1}$.
\end{abstract}

\keywords{catalogs --- surveys --- galaxies: clusters: general --- 
galaxies: distances and redshifts --- cosmology: observations --- 
large-scale structure}

\section{Introduction}

Shapley (1930) was the first to note a ``remote cloud  of galaxies'' with a
``great linear dimension, ... numerous population, and  distinctly elongated
form'' at $\alpha =$ 13\hh 30\mm,  $\delta =$ -30\deg in the direction of
Centaurus and Hydra. This cloud, commonly known today as the Shapley
Supercluster (hereafter SSC), is located at a distance of  $\sim 140 h^{-1}$
Mpc (taking the Hubble constant as  $H_{0}=100 h$ km s$^{-1}$ Mpc$^{-1}$),
beyond the Hydra-Centaurus
supercluster (at $\sim 40 h^{-1}$ Mpc).  The SSC appears to be the most
massive concentration of  matter at $z<0.1$ (Scaramella et al. 1991;
Raychaudhuri 1989, 1991; Zucca et al. 1993; Einasto et al. 1997; but see also
Batuski et al. 1999 for evidence for other, comparable structures). Over the
last 20 years, several efforts have been underway to collect data at different
wavelength bands, in order to map the matter distribution  and evaluate the
dynamical state of the SSC. A first effort in this direction was the work of
Melnick \& Quintana (1981), as a part of a program to study the physical
properties of x-ray clusters.  This study was extended to cover a significant 
fraction of the clusters in the area to estimate the masses of the member
clusters and trace the mass distribution in the supercluster,  as first
discussed by Melnick \& Moles (1987) using data reported by Quintana et al.
(1995).  The central region (around the clusters A 3558, A 3556, and A 3562) 
and other major
clusters have also been studied optically by Cristiani et al. (1987),
Metcalfe et al. (1987), Teague et
al. (1990), and recently by Bardelli and collaborators  (Bardelli et al.
1994, 1998, 2000). Several papers have investigated the X-ray emission of
clusters in the general area (Raychaudhury et al. 1991; Breen et al. 1994;
Ettori et al. 1997) and particularly in the core region of the SSC (Kull \&
B\"ohringer 1999; Hanami et al. 1999)

The SSC is a high-density region comprising  more than 30
Abell clusters of galaxies (Abell et al. 1989, hereafter  referred to as ACO)
within a radius of $\sim 25 h^{-1}$ Mpc. The central core is dominated by ACO
clusters A 3556, A 3558, and A 3562 and two poor clusters.  Two other main
concentrations are present, formed by ACO clusters A 3528/3530/3532 and
A 3570/3571/3572/3575.

This is Paper II in a series on spectroscopic observations of the Shapley
supercluster region initiated by Quintana et al. (1995; hereafter  Paper I). 
Paper I discussed spectroscopic observations in the central region of  the SSC,
and used them to estimate its mass and its effect on the motion of the Local
Group. Here, we present a significantly extended sample, including further
radial velocities of galaxies observed on a wide area which extends
approximately from 12\hh 53\mm to 13\hh46\MM5 in right  ascension and -28\deg
to -34\deg in declination, and thus covers 11.5\deg$\times$ 6\deg ,  or
$\sim 26 \times 14 h^{-2}$ Mpc$^{2}$ at the mean redshift of the SSC ($z \sim
0.05$). These data are combined with redshifts from the literature, covering
(more sparsely and irregularly) the area 12\hh 40\mm to 14\hh 10\mm in right 
ascension and
-23\deg to -39\deg  in declination (19\deg $\times$ 16\deg or $45 \times 37
h^{-2}$ Mpc$^{2}$) to give a total sample of 5090 galaxy redshifts. These are
used for a qualitative study of the general shape of the distribution  of
galaxies in the SSC, which includes filaments and sheets.  In Paper III
(Reisenegger et al. 2000)  we discuss the global supercluster dynamics and show
that the central region  is bound and currently collapsing. This has
implications for the spatial  interpretation of the structure in the central
regions of the SSC,  which we discuss below. We leave for Paper IV (Carrasco et
al. 2000) the  discussion of the dynamics and other properties of the
individual clusters of galaxies. 

\section{Observations}

The spectroscopic observations were carried out using the fiber spectrograph 
mounted on the 2.5 m (100'') DuPont telescope at Las Campanas Observatory
(LCO), in  several observing sessions. The weather conditions in all sessions
were good, with relatively good seeing and dark sky. The multifiber system
consists of a plug plate at the focal plane, with 128  fibers running to a
spectrograph coupled to the  2D-Frutti detector (Shectman
1989).  105 to 112 fibers were used for objects.  Sixteen sky fibers were set
aside, spaced at intervals of one every 6 fibers  along the spectrograph
entrance, and positioned in a random pattern in the  plug plate. Standard
quartz lamp exposures of a white spot inside  the dome were used to correct 
for pixel-to-pixel variations of the detector. The grating angle was 
changed to several values during these long  exposures in order to properly
illuminate  the whole detector surface. Five-minute exposures with helium-neon
and  thorium-argon hollow-cathode comparison lamps were taken for wavelength 
calibration before and after each exposure. The resulting 2D-Frutti images 
have a $2048 \times 1520$ pixel area. The fiber images are $\sim 8$ pixels 
wide, separated by $\sim 12$ pixels from center to center.  Thirty-four fields
of 1.5\deg $\times$ 1.5\deg were observed in  seven sessions, covering a
combined area of  11.5\deg$\times$ 6\deg.  The exposure times were adjusted
between 60 and 180 minutes, depending  on the brightness of the galaxies and
the available observing time. Table 1 gives a schedule of our seven observing 
sessions. A total of 2868
galaxy spectra were obtained, corresponding to 2627 different galaxies whose
two-dimensional distribution is shown in Figure 1.  The observations were made
with some overlap, except for fields  383-G and 383-H. 

The positions of the observed galaxies  were taken from the Digitized Sky
Survey (DSS).  The galaxy selection was carried out on the DSS images displayed
on a computer screen, giving priority to elliptical and bright galaxies, but
avoiding what appeared to be obvious foreground objects. For each fiber field a
total of 120 candidate target galaxies were chosen by eye. The fiber assignment
program  discarded close objects (within 30'' of each other) or galaxies that
fell on fiber holes from other fields that were drilled on the same metallic
plug plate. The low priority assigned to spiral or irregular-looking objects
and the lack of an objective photometric catalog imply that extreme care must
be exercised for any statistically oriented analysis of our velocity sample. It
does not have a well-defined magnitude cut-off and has a strong bias in
morphological selection. 

Gaps and areas remaining to complete a full rectangular area down to
$\delta=-35.5$\deg have been observed later. The full data set will be
published once the reduction of these remaining data is complete.

\section{Reductions}

All reductions were performed inside the  IRAF\footnote[1]{IRAF is distributed
by NOAO, which is operated by the  Association of Universities for Research in
Astronomy  Inc., under contract with the National Science Foundation.}
environment.  Velocity determinations were carried out using the
cross-correlation technique  and identifying and fitting line profiles by eye.
A complete discussion of the reduction process is given in Quintana et al.
(1996).  Here we give a summary of the reductions.  The fiber$+$2D-Frutti
system presents an S-shaped distortion of the spectra, inherent to this
instrument. A sixth-order cubic spline was fitted to trace this shape. The IRAF
Hydra package was used to extract spectra from the two-dimensional array,
correct pixel-to-pixel  variations via dome flats, use a fiber transmission
table for appropriate  sky subtraction, and  put the spectra on a linear
wavelength scale. The eighteen sky spectra  from each  exposure were combined
via a median filter and subtracted from each of  the object  spectra. The final
sky-subtracted spectra are based on the relative flux  normalization of the 
5577\AA, 5890\AA, and 6300\AA\ night-sky lines. The wavelength solutions  for
20-30  points using a 4th or 5th-order Chebyshev polynomial typically yielded
rms  residual values $< 0.5$ \AA.  With a 600 line mm$^{-1}$ grating blazed at
5000\AA, the final galaxy spectra  have a resolution of $\sim 10$\AA, a pixel
scale of $\sim 2.6$\AA\ and a  wavelength range of $\sim 3500-6900$\AA.

The radial velocities of the galaxies were determined by two different 
methods. For  non-early-type spectra (i.e. emission lines, E$+$A, etc.), a
line-by-line  Gaussian fit was used and the resulting velocities from each line
fit were averaged.  For normal  early-type spectra, the XCSAO cross-correlation
algorithm in the IRAF  RVSAO package  (Kurtz et al. 1992) was used (this
algorithm is described in Tonry \& Davis  1979). Four  template spectra with a
high signal-to-noise ratio were used to correlate  with the observed spectra in
order to determine the radial velocities of the galaxies.  Two of the {\em
templates} used here were galaxy spectra taken with the fiber  instrument 
(NGC 1407 and NGC 1426). Another template, galaxy NGC 1700, was from the  previous
detector (Shectograph) mounted on the same telescope, and one was synthetic.
The synthetic template was constructed from the stellar  spectra library of
Jacoby et al. (1984), using the ratios of stellar light for the E0 galaxy
NGC 1374 from the synthesis studies of Pickles (1985). The $R$ value (see Tonry
\& Davis 1979 for details) was used as a reliability factor of the quality of
the  measured velocity. For $R>4$, the template which gave the lowest error
value, out of the four radial velocity cross-correlation templates mentioned
above, is used as resulting velocity. (This method gave
more  consistent results than other combinations we tried, such as a
mean of the results of the four correlations.)  For $R\le 4$, we looked at the
spectra and tried line-by-line  Gaussian fitting. Finally, the radial
velocities were corrected to the  heliocentric reference frame.

\section{Building the velocity catalog}

In order to generate a large catalog of galaxy velocities to be used in the
analysis below and in further studies to be presented in Papers III and IV, the
2868 velocity determinations presented above were combined with additional data
previously published by us or obtained by other authors. In order to cover
enough area, we searched the literature  for velocities
between  12\hh 40\mm $<$ $\alpha$ $<$ 14\hh 10\mm, and  -39\deg $<$ $\delta$
$<$ -23\deg. 
Table 2 lists the references, including an unpublished set kindly provided by
D. Proust (private communication).  Many galaxies have been observed more than
once (both in different data sets or within our own observations), which allows
us both to obtain more accurate velocity determinations for these galaxies and
to  compare different data sets for general quality of the data and possible 
zero-point shifts. Below, we describe how we combine the different data sets to
build a coherent catalog.

\subsection{Comparison}

First, we compare the velocities of 241 galaxies observed by us more than  once
in different sessions. Figure 2 shows the residuals of the velocities  measured
with fibers as a function of the internal error for these  galaxies. A
zero-order polynomial fitting of the data with a $3\sigma$ clip  gives a mean
residual of $\sim 16$ km s$^{-1}$ with an rms of  $\sim 70$ km s$^{-1}$, which
is consistent with the average external errors  of $\sim 70$ km s$^{-1}$ for
the Las Campanas Redshift Survey (Shectman et al. 1996).  However, if we assume
that the quoted errors for the galaxies are good  estimates of the
cross-correlation errors, we can compute the mean and the rms of 
$(v_{fib1}-v_{fib2})/\sqrt{\sigma_{fib1}^{2}+\sigma_{fib2}^{2}}$ in order to
estimate the  true statistical errors of the data. We obtain a mean of $-0.31$
and rms of $1.57$. These  results show that the velocity zero-point correction
is negligible, and that the  cross-correlation errors are, on average, 1.57
times smaller than the true errors.  Similar correction factors were obtained 
by Malamuth
et al. (1992) (1.6) and Bardelli et al. (1994) (1.87).  As final velocity
errors, we adopt the cross-correlation errors multiplied by 1.57,
unless this result turns out to be less than 50 km s$^{-1}$. In this case, we
take the latter constant value as the velocity uncertainty of  the galaxy, as
any smaller values are likely to be statistical artifacts.

In order to combine our data with those from the  literature, we normalized all
radial velocity sets to a common zero-point. If the zero-points of different
instruments differed by significant amounts, we could introduce serious
systematic errors in the  resulting velocity dispersion and in the dynamical
analysis. Table 2  summarizes the results of the zero-point shifts obtained
from comparison of our data with the previously published  ones, totaling 919 
velocities, where, for each set, we present the number of galaxies in  common
with the present fiber  observations. A large set of data came from our previous work 
(Paper I: codes Q5, Q7), using the same telescope, the LCO
spectrograph with the same grating, and similar detectors. Therefore, one should expect
small systematic differences between these data sets.  Figure 3 shows the
residual for 65 galaxies in common with Paper I  (data sets Q5 and Q7). The
mean difference between both data sets  (8 km s$^{-1}$, with an rms of 83 km
s$^{-1}$) is negligible and well within  individual errors. It is also
consistent with fiber-to-fiber comparisons.  This confirms the consistency
between the fiber and Shectograph data,  allowing both data sets to be treated
as one. We have applied the zero-point  shift given in Table 2 to all
velocities of the literature  before combining them into a final average value.
For references without velocities in common with our data, or when the number
of velocities was  too low to obtain a meaningful shift (for example, Dressler
\& Shectman 1988),  we compared the data with other data sets in the
literature, calculating the  zero-point shift in a transitive way.

\subsection{Final Average Velocities}

The average velocity for galaxies with multiple references was calculated
following Quintana et al. (1996). This final velocity is a weighted mean 
velocity given 
by  

\begin{equation}
\overline{V} = \frac{\sum_{i} w_{i}v_{i}}{\sum_{i} w_{i}}
\label{eq1}
\end{equation}

\noindent where $v_{i}$ is the galaxy velocity, $w_{i}=1/\sigma_{i}^{2}$ is the
weighting factor, and $\sigma_{i}$ is the uncertainty. The weighting factor
$w_{i}$ came  from our  internal uncertainties or from the published
uncertainties. Our internal  uncertainties were obtained from the XCSAO
cross-correlation program as $\sigma_{int}=0.75W/(1+R)$, where $R$ is the
Tonry-Davis number and $W$ is the FWHM of the  parabola used to fit the
correlation peak, corrected by 1.57 as discussed in \S 4.1. 

Different criteria were used to calculate the error of the weighted mean
velocity.
These criteria depend upon whether we considered all velocities to be of
comparable quality or not (beyond their formal errors):

\begin{enumerate}
\item If we assume that all velocities are fully believable, then we use 

\begin{equation}
\sigma_{\mu err} = \frac{1}{\sqrt{\sum_{i} w_{i}}}
\label{eq2}
\end{equation}

as the error of the weighted mean.
\item If we assumed the quoted velocity errors to be only weights and not to
represent the true velocity errors, the uncertainty of the weighted mean is
given by the weighted {\em rms} dispersion

\begin{equation}
\sigma_{\mu W} = \sqrt{\left(\frac{\sum_{i} w_{i} v_{i}^{2}}{\sum_{i} w_{i}} 
- \overline{V}^{2}\right)\times \frac{1}{N-1}}
\label{eq3}
\end{equation}

\item If internal uncertainties adequately measure the external uncertainties,
and if the number of the individual measurements is large, both ways should
give the same results in the mean, but $\sigma_{\mu err}$ is too low if
internal uncertainties underestimate external ones, and $\sigma_{\mu W}$ is too
low if by chance the  measured values are close to each other. To detect
whether the differences between  measured values are clearly larger than the
internal uncertainties or not, we compared the unweighted {\em rms} dispersion
$\sigma_{UW} = \sqrt{\sum_{i} (v_{i} -  \overline{V})^{2}/N(N-1)}$ to
$\sigma_{T} = \sqrt{\sum_{i} \sigma_{i}/(N-1)}$, as a criterion to estimated
the error of the weighted mean velocity. The criterion is the following: {\it
i)} if $\sigma_{UW} > 2 \sigma_{T}$, we don't have any information about which
measurement is correct (systematic errors should be blamed), so the uncertainty
of the weighted mean is given by $\sigma_{UW}$, without dividing by 
$\sqrt{N}$; {\it ii)} if the anomalous case detected had values of $\sigma_{\mu
W}$  and $\sigma_{T}$ lower than 50 km s$^{-1}$, we assumed $\sigma_{\mu W}$ 
(eq. 3).
\end{enumerate}

In a few cases we have discarded a velocity when analysis of the spectra, the galaxy 
morphology, or other details show discrepancies which are large compared 
to the measurement error. Velocities reported in the literature which differed
by $>200$ km s$^{-1}$ from our own determination for the same galaxy
were discarded before combining. 

The final sample resulting from this procedure contains 5090 galaxies with
known redshifts. Of these, 2627 were observed by us, including 567 in
common with the literature, and 2060  for which this is the first velocity
determination. 
Therefore, the present study contributes 41\% of the presently available sample
of galaxies in the region (and an even higher fraction in the restricted region
covered by this study and in the redshift range corresponding to the SSC).

\section{Morphology}

The full sample of galaxies to be used in the present study and in those 
following it (Papers III and IV) is presented in Figures 4 - 7.  Their spatial
distribution, shown in Figure 4, makes it clear that the area, particularly
outside our original survey, is covered in a very sparse and patchy way. In
addition, it has to be stressed that, even in the areas covered by the surveys,
this sample does not have a well-defined completeness limit\footnote[2]{Based
on the COSMOS/UKST galaxy catalog of the southern sky  (Yentis et al. 1992), we
estimate  the completeness of our own data, in the regions covered, to be 
$\sim 75$\% down to $b_{j} < 18.5$ mag.}. This has to be kept in mind and 
should deter
us  from overinterpreting the apparent structures. On the other hand, the level
of detail visible in all these figures makes it clear that a more detailed
assessment of the complex morphology of this region, particularly of the SSC
itself, should now be possible. 

The histogram in Figure 5 presents four maxima. The main body of the SSC is
represented by  the highest peak, which is centered at $\sim$ 15,000 
km s$^{-1}$ and
extends over the  velocity range 13,000 - 18,000 km s$^{-1}$. Another peak 
between 9,000 and 13,000 km s$^{-1}$ shows a foreground structure connected to
the SSC.  This can be seen better in the cone diagram of Figure 6 (right
ascension projection). The other two peaks  are at 4,000 km s$^{-1}$ (the
Hydra-Centaurus connection and the ``Great Attractor'' region) and at 23,000
km s$^{-1}$ (another possible supercluster behind the SSC). Figures 5 and 7 
also show a few ``clumps'' at higher redshift. The cone
diagram confirms the presence of a large void  in the velocity interval 18,000
- 22,000 km s$^{-1}$ behind the SSC (cf. Bardelli et al. 2000). 

\subsection{The Shapley Supercluster}

The definition of the extension of the Shapley Supercluster is not an easy 
task, because of the complexity of the structures in the velocity  distribution
and the limited extension of the velocity survey on the plane  of the sky. The
presence of many clusters, with their characteristic finger-of-God velocity
structures, complicates the study. Moreover, 
the remaining irregularities and gaps in the observations
can mimic apparent structures. Finally, as today's redshift surveys show, 
dense structures are linked by
filaments and walls, forming a fabric that completely fills the space.
Therefore, we can expect that the SSC will be linked to foreground and
background structures, as well as others on the plane of the sky. 

Where does
the SSC proper end? Several criteria can be used as definitions for its extent,
such as cluster or galaxy density or density contrast, percolation criteria,
and others. Without a redshift survey over a wider area, we feel we cannot
address this question more reliably than has been done before (e.g., Zucca et al.
1993; Einasto et al. 1997). As a preliminary criterion, based  primarily on the
velocity histogram of Figure 5, we will consider as SSC members all galaxies
and clusters in the area surveyed that have a redshift 9,000 $\leq v\leq$
18,000 km s$^{-1}$, corresponding to a total depth of $90 h^{-1}$ Mpc. 
These limits are supported by the cone diagrams, which also
clearly show voids of much lower density, both in front and behind this
structure. In this redshift range, there are 
2949 redshifts in the total sample, allowing a fairly detailed study of the
SSC's structure.

Because of the limited sky coverage of the redshift surveys, the transverse 
dimensions of the SSC are less clear, but it appears to cover much of the 
surveyed area, or at least $35 \times 25 h^{-2}$ Mpc$^{2}$. In this volume,
quite elongated along the line of sight, the central region of the SSC, 
centered on the rich cluster A 3558, is quite 
outstanding. Several filaments and sheets of galaxies are also apparent, 
extending at different densities between the central region and surrounding 
groups of clusters, as detailed below. Galaxy clusters in this region include
many ACO clusters, in addition to other groups and clusters that have been
identified, particularly from X-ray surveys. Among these are the  poor clusters
SC 1327-312 and SC 1329-314 (Breen et al. 1994) and the ROSAT  cluster RX
1252.5-3116 (Pierre et al. 1994), which have recently been found  through x-ray
and optical observations, and the groups SC 1336-314, SC 1340-294, and 
SC 1342-302
found in the present study  (see Paper IV for more details). The full list of
clusters in the surveyed region is given in Table 3, and their positions are 
shown on Figure 4. 

Before entering into the detailed discussion of the supercluster's morphology,
we want to make an important point. In dense structures such as the SSC, the
internal gravity can produce strong peculiar motions, which implies that 
relative redshifts are not an exact  representation of relative distances,
especially not on small scales. In particular, virialized clusters, roughly
spherical in real space, produce the characteristic ``finger-of-God'' effect in
redshift space, but also somewhat larger structures can be strongly affected.
It is argued in Paper III
that the central part of the SSC (within a radius $\sim 10 h^{-1}$ Mpc of
the main cluster, A 3558) is presently collapsing, with infall velocities $\sim
1000$  km s$^{-1}$. Therefore, objects within this region, with a
somewhat {\it  larger redshift} than A 3558, are probably {\it closer} to us
than A 3558 and  falling towards it from the front side, while objects with
somewhat {\it  lower redshifts} are likely to be {\it further away.} The model
described in Paper III is too crude to be expected to account for any details
of the observed structures or to transform reliably from redshift space to real
space, therefore the discussion below will be done {\it as if} redshift truly
represented relative distance. This is a good approximation for redshift  differences of
several thousand km s$^{-1}$, but it should be kept in mind that it is unlikely
to be correct for much smaller differences.

In the detailed study of the SSC's substructures, we are guided by the wedge
diagrams in Figure 6  
and by many plots of the galaxy and cluster positions in velocity cuts of
different depths, the most illuminating of which are presented in Figure 8. 
Although we obtain fairly clear identifications, we
have to keep  in mind the uncertainties introduced by the mentioned gaps and
irregularities  remaining in the velocity survey, so that our results are
provisional and  require confirmation when we complete the survey.

The {\it Central Region} of the SSC (CR), roughly spherical in shape, 
has at its core the highest-density, elongated region containing the
Abell clusters A 3562, A 3558, and A 3556 (Fig. 8f), with almost
identical recession velocities near 14400 km s$^{-1}$, and the groups
SC 1329-314 (Fig. 8e) and SC 1327-312 (Fig. 8h),
whose more discrepant velocities (by several hundred km s$^{-1}$) could be 
attributed to infall along the line of sight. To the south of the elongated 
feature, the CR also contains the cluster A 3560 (Fig. 8g). Note that the 
very thin velocity slice of Figure 8f (chosen to eliminate other structures) 
does not do
justice to the extreme galaxy richness of this region. Given its high velocity
dispersion, the CR can be seen over Figures 8d through 8i. The whole of this
central region and all of its immediate surroundings are within the volume
which, according to the  study of Paper III, is currently undergoing
gravitational collapse. Therefore, in this volume, {\it redshift should
anticorrelate with distance}.

From this central region, a wide wall of galaxies, groups and clusters, 
which we hereafter call the {\it Front Eastern Wall} (FEW), extends to the east and 
front of the supercluster at $\sim$ 10,000 - 11,000 km s$^{-1}$ in redshift 
(Fig. 8c). The densest part of 
this wall is a filament running eastward at a declination just south 
of the CR and containing the newly identified group SC 1336-314 and the
clusters A 3571/3572/3575. A 3570 is located at the southern tip of the
observed part of the FEW, and A 3578 may mark its northern end, which is less
well sampled. The northern part of the FEW has rather low galaxy density,
but is wider (along the line of sight) than the southern part and connects 
to the cluster A 1736a and
another, unidentified group just south-east of it (Fig. 8b). From these, 
a filament runs to the south-west, connecting to A 3537a (Fig. 8a). 

Another major feature is the wide, dense {\it North-West Filament} (NWF) that
from the CR runs first SW and then turns to the NW, going behind and  away from
the CR. It includes A 3554, A 726S, A 724S, and A 3544 in the SW 
(Figs. 8f-g), A 3537b further west 
(Fig. 8h), and finally the group
of rich X-ray clusters A 3528/3530/3532, and RX J1252.5-3116, 
all in a narrow redshift range  and prominently visible on Fig. 8i, at
the furthest point of the SSC, both in redshift and in angular position. Also
associated with this filament may be  the clusters A 3552,\footnote[3]{The
existence of A 3552 as a true cluster is  doubtful: No X-rays have been detected
from it, and  at its position only a small group of galaxies can be found
concentrated in redshift. Its
previous, large mass measurements (e.g., Paper I) are likely an artifact of
having  included a number of ``field'' galaxies which, as they belonged to the
SSC,  showed a smoothly peaked, but very extended redshift distribution.}
A 3553, A 731S, A 729S, and A 718S (Figs. 8g-h),
which do not follow the filamentary structure, but
are spatially close to it and in the same redshift range. The specific  initial
direction and later turn of the NWF observed in the data may be  an artifact of
the gap in observed galaxies in this region, and are certainly  affected by the
internal dynamics (peculiar velocities) in the SSC.

The last major feature of the SSC is the {\it Northern Extension,} a broad arm 
running N from the CR and to progressively lower redshift (Fig. 8e),
connecting to the clusters A 3557a ($v\approx$ 14,300 km s$^{-1}$; Fig. 8f), 
A 3559 and A 3555 ($v\approx$ 14,000 km s$^{-1}$),
A 1736b ($v\approx$ 13,600 km s$^{-1}$), and beyond.

There are two other minor arms or filaments extending out from the CR:

\begin{itemize}
\item[a)] Behind the FEW and at slightly higher redshift than the CR (Fig.
8g),  a diffuse filament from the CR to the North-East  leads through the new
cluster SC 1342-302 to A 3577.

\item[b)] From the CR to the SE and to higher redshift (Fig. 8h), a broad arm 
connects to the group of clusters A 3564, A 3566, A 3568, and A 742S  (all with
$v\approx$ 15,200 - 15,500 km s$^{-1}$). This arm shows a dense cloud  just
north of the catalog position of A 3568, which may be part of this cluster (if
its catalog center position  were somewhat off), or an artifact of the galaxy
selection at the edges of the fiber plates.
\end{itemize}

\subsection{Background structures}

On the high-redshift side of the central concentration, a large, very low
density void, also identified by Bardelli et al. (2000)  as {\it V1,} opens up
and extends from $\sim$ 18,000 to 21,000 km s$^{-1}$ (see Figs. 5, 6, 7, and 8j).
The only two clusters on Fig. 8j are near the upper end of this redshift range,
and at the south and west edges of the studied area.

Behind the void, there appears a low-density, transverse, S-shaped filament at 
$v\approx$ 21,300 km s$^{-1}$, extending diagonally across the surveyed area
(Fig. 8k).
Possibly connected to it is the most prominent of all structures appearing 
behind the SSC. This structure appears as a strong peak 
around 23,000 km s$^{-1}$ in the velocity histogram (Fig. 5), and can be seen in more
detail in Figure 8l. It extends only $\sim$ 2000 km s$^{-1}$ in redshift and 
$\sim 25 h^{-1}$ Mpc in its larger transverse dimension, but contains six Abell 
clusters (in order of increasing redshift, A 3545/3524/3531/3549/1648/3557b),
in addition to the new cluster, SC 1340-294, and another candidate cluster 
near $12^h55^m$ in right ascension and -29\deg in declination. In the better
sampled area in the center of Figure 8l, it can be seen that the clusters are
also connected by abundant intercluster galaxies. Thus, we believe that this
concentration is most aptly described as another supercluster.

Beyond this redshift, the coverage of our sample is not much better than the
one studied by Bardelli et al. (2000), which has been included.  For this
reason, we do not go into details of its analysis. Fig. 7 shows essentially the
same structures (concentrations and voids) that have been identified by
Bardelli and collaborators. The possible  superclusters S300a and S300b are
clearly present (both are at right ascension $\sim 13^h10^m$, at redshifts
$\approx$ 32,000 and 40,000 km s$^{-1}$, respectively), but not much structure
can be seen within them, although hints of some clusters of galaxies are
apparent in S300a, which will be further studied in Paper IV. We only add that
both candidate superclusters seem to be connected by a bubble-like structure 
surrounding the void V3 and including on its surface the cluster A 3551 
($v\approx$ 37,500 km s$^{-1}$).

\section{Conclusions}

The large number of new redshift determinations in the Shapley Supercluster
region brings the total number of available galaxy redshifts in the SSC
to 2949. For comparison, the next best-sampled superclusters are
Corona Borealis, for which this number is 528 (Small et al. 1997), and the
Hercules Supercluster, with 468 available velocities  (Barmby \& Huchra 1998).
This allows to analyze the morphology of this supercluster in much more detail
than is possible elsewhere. We show that several structures extend outward from
the supercluster core, linking it to groups of galaxy clusters up to $\sim 40
h^{-1}$ Mpc away. We also find a few new clusters or groups in the SSC region,
and identify another supercluster in the background at $\sim$ 23,000 km
s$^{-1}$. 

\acknowledgements

We are grateful to D. Proust for making his redshift data available to us prior
to publication. We also thank the Director of the Observatories of the Carnegie
Institution of Washington for generous allocation of telescopes at Las Campanas
Observatory. This work benefitted from the use of the
NASA/IPAC Extragalactic Database (NED), which is operated by the Jet Propulsion
Laboratory, California Institute of Technology, under contract with the
National Aeronautics and Space Administration. It was financially supported by 
FONDECYT grant 8970009 ({\it Proyecto de L\'\i neas Complementarias}), and by 
a Presidential  Chair in Science awarded to HQ. ERC was funded by FAPESP PhD
fellowship 96/04246-7.

\newpage

\figcaption[fig1.eps]{Two-dimensional distribution of the 2627 galaxies 
observed with the fiber spectrograph. The solid squares indicate the 34 
observed fields. Each field covers $1.5^{\circ} \times 1.5^{\circ}$. 
\label{fig1}}

\figcaption[fig2.eps]{Residual of the heliocentric velocities as a function of
the internal quadratic error for the 241 galaxies observed  with the fiber
spectrograph in different sessions. The solid line represents  the zeroth-order
polynomial fit. The average shift after 3$\sigma$ clipping  is $-13$ km
s$^{-1}$, with a rms of $65$ km s$^{-1}$. The cross-correlation  errors are, on
average, 1.57 times smaller than the true statistical errors.  \label{fig2}}

\figcaption[fig3.eps]{Residual of the heliocentric velocity as a function of 
the internal quadratic error for the 65 galaxies in common with our previously 
published data (Q5, Q7 in Paper I). The mean shift of $8$ km
s$^{-1}$ is  negligible compared to the $83$ km s$^{-1}$ rms of the residuals,
and  consistent with the fiber-to-fiber comparison. \label{fig3}}

\figcaption[fig4.eps]{Both panels show the two-dimensional distribution of 
the full sample of galaxies studied in
the present paper, including new data and data from the literature between 
12\hh 40\mm $<$ $\alpha$ $<$ 14\hh 10\mm, and  -39\deg $<$ $\delta$ $<$
-23\deg. The open circles represent the positions of the clusters of
galaxies given in Table 3. Panel (a) shows the clusters with known redshifts
in the range of the SSC, 9,000 $< v <$ 18,000 km s$^{-1}$, with the sizes
of the circles representing the Abell radius, $1.5 h^{-1}$ Mpc, at the 
distance inferred from their individual redshifts. Panel (b) shows all other 
clusters, i.e., all those whose redshifts are either unknown
or outside the SSC range. In this case, the size of the circles was taken 
inversely proportional to their ACO distance class. Clusters without a
distance class are shown as crossed circles.\label{fig4}}

\figcaption[fig5.eps]{Radial velocity distribution for galaxies observed by us 
(dashed histogram) and for the total sample including data from the literature
(open histogram), with $v <$ 50,000 km s$^{-1}$.\label{fig5}}

\figcaption[fig6.eps]{Cone diagrams of all galaxies with known redshifts in the
region limited by 12\hh 50\mm $<$ $\alpha$ $<$ 14\hh 05\mm, -38\deg $<$
$\delta$ $<$ -24\deg, and 6,000 $< v <$ 20,000 km s$^{-1}$. Panel (a) shows
right ascension vs. radial velocity; panel (b) shows declination
vs. radial velocity.\label{fig6}}

\figcaption[fig7.eps]{Cone diagrams of all galaxies with known redshifts in the
region limited by 12\hh 50\mm $<$ $\alpha$ $<$ 14\hh 05\mm, -38\deg $<$
$\delta$ $<$ -24\deg, and 18,000 $< v <$ 50,000 km s$^{-1}$. Panel (a) shows
right ascension vs. radial velocity; panel (b) shows declination
vs. radial velocity.\label{fig7}}

\figcaption[fig8a.eps,fig8b.eps]{Distribution on the sky of galaxies and
clusters with known redshifts in thin velocity slices of various widths,
illustrating the three-dimensional structure of the SSC and a background
supercluster:  (a) 9,000 $< v <$ 10,000 km s$^{-1}$; (b) 10,000 $< v <$ 11,000
km s$^{-1}$;  (c) 11,000 $< v <$ 12,000 km s$^{-1}$ {\it (Front Eastern Wall)}; 
(d) 12,000 $< v <$ 13,000 km s$^{-1}$;  (e) 13,000 $< v <$ 14,100 km s$^{-1}$
{\it (Northern Extension)};  (f) 14,100 $< v <$ 14,500 km s$^{-1}$ {\it (Central
Region)};  (g) 14,500 $< v <$ 15,000 km s$^{-1}$; (h) 15,000 $< v <$ 16,000 km
s$^{-1}$;  (i) 16,000 $< v <$ 18,000 km s$^{-1}$ (end of {\it NW Filament});  (j)
18,000 $< v <$ 21,000 km s$^{-1}$ (void {\it V1});  (k) 21,000 $< v <$ 22,000 km
s$^{-1}$;  (l) 22,000 $< v <$ 24,000 km s$^{-1}$ (background supercluster).
\label{fig8}}

\newpage
\begin{deluxetable}{clccccc}
\footnotesize
\tablecolumns{7}
\tablewidth{0pt}
\tablecaption{Observing sessions \label {tab:tab1}}
\tablehead{
\colhead{Session} & \colhead{SRC/ESO plate} & \colhead{Field id.} &
\colhead{$\alpha_{2000}$} & \colhead{$\delta_{2000}$} & 
\colhead{$t_{exp}$} & \colhead{Code} \\ 
\colhead{} & \colhead{} &
\colhead{} & \colhead{$^{h}$ $^{m}$ $^{s}$} & 
\colhead{$^{\circ}$ $^{'}$ $^{''}$ } &\colhead{s} & \colhead{}}
\startdata
 01/1992 & 444                & 444-A & 13 23 50.6 & -29 32 44 & 5400 & QC01 \\
         & 442/443$^{a}$      & 443-A & 12 56 03.1 & -29 32 01 & 5400 &      \\
 05/1993 & 444$^{a,b}$        & 444-B & 13 31 09.0 & -30 44 25 & 5400 & QC02 \\
         & 444$^{a,c}$        & 444-C & 13 31 09.0 & -30 44 25 & 5900 &      \\
         & 444                & 444-D & 13 24 11.7 & -30 47 48 & 9000 &      \\
         & 383/444$^{a,b}$    & 383-A & 13 31 27.5 & -32 13 50 & 5400 &      \\
         & 383/444$^{a,c}$    & 383-B & 13 31 27.5 & -32 13 50 & 9000 &      \\
         & 382/383/444$^{a}$  & 383-C & 13 24 19.4 & -32 14 32 & 5400 &      \\
 03/1994 & 382/443/444$^{a}$  & 382-A & 13 18 50.1 & -32 01 46 & 6000 & QC03 \\
         & 383/444/445$^{a}$  & 383-D & 13 37 02.5 & -32 06 14 & 6600 &      \\
         & 443/444$^{a}$      & 443-B & 13 18 35.7 & -30 46 49 & 7200 &      \\
         & 444/445$^{a}$      & 444-E & 13 36 06.4 & -30 52 53 & 6600 &      \\
         & 443/444$^{a}$      & 443-C & 13 18 37.9 & -29 30 54 & 7200 &      \\
         & 444                & 444-F & 13 24 49.3 & -29 31 41 & 7200 &      \\
         & 444                & 444-G & 13 30 03.8 & -29 30 33 &10800 &      \\
         & 444/445$^{a}$      & 444-H & 13 36 55.9 & -29 33 04 & 8400 &      \\
 05/1994 & 382                & 382-B & 13 10 41.1 & -33 15 30 & 3600 & QC04 \\
         & 382                & 382-C & 13 17 18.0 & -33 15 19 & 5400 &      \\
         & 382                & 382-D & 13 23 51.8 & -33 14 56 & 5400 &      \\
         & 383                & 383-E & 13 30 24.5 & -33 19 52 & 5400 &      \\
         & 383                & 383-F & 13 36 27.7 & -33 16 47 & 3600 &      \\
         & 443                & 443-D & 13 12 39.3 & -29 32 44 & 8100 &      \\
 02/1995 & 443                & 443-E & 12 56 52.7 & -28 21 03 & 6600 & QC05 \\
         & 443                & 443-F & 12 56 51.7 & -30 44 21 & 7000 &      \\
         & 443                & 443-G & 13 03 22.4 & -30 46 02 & 5000 &      \\
 03/1995 & 383                & 383-G & 13 29 52.2 & -34 45 27 & 4800 & QC06 \\
         & 383                & 383-H & 13 42 51.6 & -34 46 46 & 5100 &      \\
         & 443                & 443-H & 13 03 04.7 & -29 16 57 & 5400 &      \\
         & 443                & 443-I & 13 05 31.7 & -32 05 48 & 5400 &      \\
         & 444                & 444-I & 13 24 48.5 & -28 27 37 & 4800 &      \\
         & 444                & 444-J & 13 30 54.2 & -28 25 44 & 4800 &      \\
 01/1996 & 445                & 445-A & 13 43 07.3 & -32 00 15 & 6600 & QC07 \\
         & 445                & 445-B & 13 42 43.1 & -30 25 14 & 7200 &      \\
         & 445                & 445-C & 13 42 57.1 & -28 57 04 & 7200 &      \\
\tablenotetext{a}{Overlapping regions in SRC/ESO plates.}
\tablenotetext{b}{Bright galaxies.}
\tablenotetext{c}{Faint galaxies.}
\tablecomments{Columns: 1. Date of observation. 2. ESO/SRC plate 
number. 3. Our internal field identification. 4. Right ascension 
of the field center. 5. Declination of the field center.
6. Exposure time. 7. Code for observing session.}
\enddata
\end{deluxetable}

\newpage

\begin{deluxetable}{lrrccc}
\footnotesize
\tablecolumns{6}
\tablewidth{0pt}
\tablecaption{Zero point shift between fibers and the data from the literature.
\label {tab:tab2}}
\tablehead{
\colhead{References} & \colhead{$N_{ref}$} & \colhead{$N_{comm}$} &
\colhead{$\Delta$V} & \colhead{$\sigma_{\Delta_{V}}$} & \colhead{rms} \\ 
\colhead{} & \colhead{}  & \colhead{}  & \colhead{km/s} & \colhead{km/s} & \colhead{km/s}}
\startdata
Bardelli et al. 1994                     & 311 &  94 &   -5 & 10 &  93 \\
Bardelli et al. 1998                     & 174 &  73 &   -7 & 12 & 108 \\
Bardelli et al. 2000                     & 442 &  90 &   -7 & 16 & 102 \\
Cristiani et al. 1987                    &  44 &  35 &   50 & 26 &  90 \\
Da Costa et al. 1986                     & 111 &  29 &  -35 & 16 &  84 \\
Da Costa et al. 1987                     & 139 &  20 &  -38 & 19 &  82 \\
Dressler 1991                            & 545 &  22 &  -39 & 19 &  84 \\
Drinkwater et. al. 1999                  & 306 &  87 &   -9 & 12 & 110 \\ 
Melnick \& Quintana 1981                 &  26 &  19 &  -65 & 73 &  93 \\
Metcalfe et al. 1987                     &  40 &  27 &   10 & 15 &  70 \\
Proust 2000 (private communication)      & 339 &  31 &    7 & 13 &  64 \\
Quintana et al. 1995 (Q5 and Q7)         & 161 &  65 &    8 & 10 &  83 \\
Quintana et al. 1995 (Q1 and Q2)         &  32 &  15 &   67 & 30 & 107 \\
Quintana et al. 1995 (Q3,Q4,Q6,M1,M2,M3) & 178 &  87 &    9 & 14 & 117 \\
Quintana et al. 1997                     & 298 &  43 &   27 & 14 &  74 \\
Richter 1987                             &  49 &  28 &  -43 & 20 &  94 \\
Stein 1996                               & 335 & 112 &  -17 &  5 &  46 \\
Teague et al. 1990                       &  90 &  25 &   69 & 17 &  77 \\
Vettolani et al. 1990                    &  42 &  17 &    4 & 19 &  68 \\
\tablecomments{Columns: 1. Reference. 2. Total number of galaxy redshits. 
3. Number of galaxies in common with the present work. 4. Mean velocity
difference (zero-point shift) with respect to our work for galaxies in common. 
5. Error in the zero-point shift. 6. Rms velocity difference for galaxies
in common. We have no velocities in common with Allen et al. (1991),
Dressler \& Shectman (1988), and Quintana \& de Souza (1993). For these cases, the
zero-point shift was calculated in a transitive way.}
\enddata
\end{deluxetable}

\newpage

\begin{deluxetable}{lcclcrc}
\footnotesize
\tablecolumns{7}
\tablewidth{0pt}
\tablecaption{Clusters of galaxies in the region 12\hh 40\mm $<$ $\alpha$ $<$ 
14\hh 10\mm, -39\deg $<$ $\delta$ $<$ -23\deg. \label{tab:tab3}}
\tablehead{
\colhead{Name} &
\colhead{$\alpha_{2000}$} &
\colhead{$\delta_{2000}$} &
\colhead{BM} &
\colhead{$D$} &
\colhead{$V$} &
\colhead{Reference} \\
\colhead{} &
\colhead{$^{h}$ $^{m}$ $^{s}$} &
\colhead{$^{\circ}$ $^{'}$ $^{''}$} &
\colhead{} &
\colhead{} &
\colhead{[km/s]} &
\colhead{for $V$}}
\startdata
A 3524          & 12 40 05.3 & -34 13 28 & I     & 5 & 22275 &  5 \\
A 706S          & 12 41 05.3 & -33 22 27 & II-III& 6 &       &    \\
A 1604          & 12 43 57.0 & -23 06 25 &       & 6 &       &    \\
A 708S          & 12 44 23.9 & -33 25 25 & I     & 6 &       &    \\
A 3527          & 12 49 53.6 & -36 45 00 & I     & 6 &       &    \\
A 714S          & 12 51 28.9 & -26 27 18 & I     & 1 &       &    \\
VMF 98-124      & 12 52 05.4 & -29 20 46 &       &   & 50965 &  8 \\
VMF 98-125      & 12 52 11.3 & -29 14 59 &       &   &       &    \\
RX J1252.5-3116 & 12 52 30.0 & -31 16 00 &       &   & 16039 &  3 \\
A 715S          & 12 53 06.2 & -27 41 01 & I     & 5 &       &    \\
A 1633          & 12 53 59.8 & -26 23 00 &       & 6 &       &    \\
A 3528          & 12 54 22.2 & -29 00 46 & II    & 4 & 16365 &  2 \\
A 3530          & 12 55 36.0 & -30 20 51 & I-II  & 4 & 16253 &  2 \\
A 3531          & 12 57 05.4 & -32 54 59 & II    & 5 & 22454 &  1 \\
A 3532          & 12 57 21.9 & -30 21 47 & II-III& 4 & 16569 &  2 \\
A 3535          & 12 57 54.7 & -28 28 46 & III   & 5 & 19986 &  2 \\
A 717S          & 12 58 05.9 & -28 12 59 & I     & 5 &       &    \\
A 1648          & 12 58 59.6 & -26 37 00 &       & 5 & 22994 &  1 \\
A 718S          & 12 59 42.2 & -33 39 59 & I     & 4 & 14330 &  5 \\
A 3537a         & 13 01 00.6 & -32 26 00 & I-II  & 2 &  9507 &  2 \\
A 3537b         & 13 01 00.6 & -32 26 00 & I-II  & 2 & 15753 &  2 \\
A 3540          & 13 03 18.1 & -33 15 59 & II-III& 6 &       &    \\
A 1664          & 13 03 41.5 & -24 12 59 &       & 6 & 38254 &  1 \\
A 3541          & 13 03 41.5 & -24 14 59 & II-III& 6 & 50665 &  1 \\
A 721S          & 13 06 05.8 & -37 34 59 & II    & 4 & 14867 &  2 \\
A 3542a         & 13 08 42.6 & -34 32 59 & I     & 4 & 27551 &  5 \\
A 3542b         & 13 08 42.6 & -34 32 59 & I     & 4 & 39393 &  5 \\
A 3543          & 13 09 23.9 & -23 29 59 & III   & 6 &       &    \\
A 3544          & 13 11 05.6 & -32 59 00 & I-II  & 6 & 14996 &  2 \\
A 3545          & 13 11 23.5 & -34 04 00 & I-II  & 6 & 22245 &  5 \\
A 3546          & 13 13 06.6 & -29 58 00 & I     & 6 &       &    \\
A 3547          & 13 13 17.6 & -37 19 00 & II?   & 6 &       &    \\
A 724S          & 13 13 17.6 & -32 56 01 & II    & 6 & 14775 &  2 \\
A 725S          & 13 14 12.6 & -30 11 00 & II    & 6 &       &    \\
A 3549          & 13 14 23.6 & -29 26 00 & I-II  & 5 & 22634 &  5 \\
A 726S          & 13 15 11.8 & -33 38 51 & II    & 5 & 14606 &  2 \\
A 3551          & 13 18 11.9 & -30 55 00 & I-II  & 6 & 37522 &  2 \\
A 3552          & 13 18 55.2 & -31 49 03 & I     & 6 & 15598 &  2 \\
A 3553          & 13 19 14.6 & -37 10 45 & I-II  & 4 & 15110 &  4 \\
A 3554          & 13 19 32.4 & -33 29 15 & I-II  & 4 & 14494 &  2 \\
A 3555          & 13 20 20.8 & -28 53 38 & II    & 4 & 13950 &  2 \\
A 728S          & 13 20 54.2 & -27 17 59 & II-III& 5 &       &    \\
A 729S          & 13 21 29.9 & -35 47 00 & I     & 5 & 14960 &  4 \\
A 730S          & 13 22 00.2 & -26 58 00 & I-II  & 5 &       &    \\
A 731S          & 13 22 59.3 & -34 52 00 & I-II  & 4 & 15140 &  5 \\
A 3556          & 13 24 06.6 & -31 40 12 & I     & 4 & 14373 &  2 \\
A 1727          & 13 24 12.2 & -23 02 00 &       & 6 &       &    \\
A 732S          & 13 24 42.4 & -26 01 00 & II    & 6 &       &    \\
CL 1322-30      & 13 24 47.6 & -30 17 37 &       &   &  4222 &  6 \\
A 3557a         & 13 24 55.2 & -28 53 16 & I-II  & 5 & 14270 &  5 \\
A 3557b         & 13 26 11.5 & -28 47 34 & I-II  & 5 & 23084 &  5 \\
J132543.5-2944 & 13 25 43.5 & -29 44 34 &       &   &       &    \\
A 733S	       & 13 25 53.9 & -37 13 59 & II	& 5 & 20906 &  1 \\
A 1736a         & 13 26 44.3 & -27 26 22 & III	& 2 & 10443 &  2 \\
A 1736b         & 13 26 48.7 & -27 08 38 & III	& 2 & 13577 &  2 \\
A 3558	       & 13 27 56.9 & -31 29 44 & I	& 3 & 14390 &  2 \\
SC 1327-312     & 13 29 47.0 & -31 36 29 & I-II  &   & 15124 &  2 \\
A 3559	       & 13 29 51.0 & -29 30 51 & I	& 4 & 13995 &  2 \\
A 736S	       & 13 31 00.6 & -28 01 59 & I-II  & 6 &	    &	 \\
SC 1329-314     & 13 31 36.0 & -31 48 46 & I-II  &   & 13388 &  2 \\
A 3560	       & 13 32 25.3 & -33 08 12 & I	& 3 & 14541 &  2 \\
A 1757	       & 13 33 32.9 & -23 16 22 &	& 5 & 37894 &  1 \\
A 3562	       & 13 33 56.8 & -31 29 23 & I	& 3 & 14377 &  2 \\
A 3564	       & 13 34 24.0 & -35 13 00 & II	& 3 & 15188 &  2 \\
SC 1336-314     & 13 36 19.0 & -31 48 00 &	&   & 11857 &  2 \\
A 3565	       & 13 36 29.8 & -33 58 17 & I	& 1 &  3268 &  1 \\
MS 1335.2-2928  & 13 38 05.8 & -29 43 55 &	&   & 56661 &  7 \\
A 3566	       & 13 39 00.5 & -35 32 59 & II	& 4 & 15423 &  2 \\
SC 1340-294     & 13 40 00.0 & -29 46 00 &	&   & 23229 &  2 \\
A 3567	       & 13 39 48.6 & -36 26 59 & I-II  & 6 &	    &	 \\
A 3568	       & 13 41 12.5 & -34 38 00 & II-III& 6 & 15483 &  2 \\
A 1771	       & 13 42 11.6 & -26 16 00 &	& 5 & 32048 &  1 \\
A 3569	       & 13 42 41.9 & -35 45 01 & I-II  & 6 &	    &	 \\
SC 1342-302     & 13 42 45.0 & -30 20 00 &	&   & 14515 &  2 \\
A 739S	       & 13 42 54.3 & -34 57 59 & II	& 5 &	    &	 \\
A 740S	       & 13 43 30.0 & -38 10 59 & I-II  & 3 & 10073 &  1 \\
A 742S	       & 13 44 36.0 & -34 18 00 & I-II  & 4 & 15277 &  2 \\
A 3570	       & 13 46 50.7 & -37 54 58 & I-II  & 4 & 11152 &  1 \\
A 3571	       & 13 47 28.3 & -32 51 52 & I	& 4 & 11730 &  4 \\
A 744S	       & 13 47 29.3 & -32 07 59 & I	& 5 &	    &	 \\
A 3572	       & 13 48 11.9 & -33 22 00 & I-II  & 4 & 12141 &  1 \\
A 745S	       & 13 48 11.9 & -36 27 59 & III	& 6 &	    &	 \\
A 3573	       & 13 48 24.3 & -34 40 00 & I	& 6 &	    &	 \\
A 1791	       & 13 48 54.5 & -25 26 00 &	& 5 &	    &	 \\
A 3574	       & 13 49 05.4 & -30 17 47 & I	& 1 &  4227 &  1 \\
A 746S	       & 13 49 48.2 & -34 57 59 & II	& 4 &	    &	 \\
A 1794	       & 13 50 11.5 & -26 18 59 &	& 5 &	    &	 \\
A 747S	       & 13 50 59.7 & -36 06 59 & II	& 5 &	    &	 \\
A 1802	       & 13 51 24.4 & -26 41 59 &	& 5 &	    &	 \\
A 3575	       & 13 52 35.9 & -32 51 59 & II	& 4 & 11242 &  5 \\
A 748S	       & 13 52 35.9 & -32 23 00 & II	& 6 &	    &	 \\
A 3576	       & 13 52 48.3 & -30 16 59 & I	& 6 & 44669 &  1 \\
A 749S	       & 13 52 59.3 & -34 20 59 & II:	& 6 &	    &	 \\
A 750S	       & 13 53 36.4 & -38 19 59 & I-II  & 6 &	    &	 \\
A 751S	       & 13 54 05.3 & -25 42 59 & I-II  & 6 &	    &	 \\
A 3577	       & 13 54 17.7 & -27 49 59 & II	& 4 & 14870 &  5 \\
A 1816	       & 13 55 41.6 & -26 20 59 &	& 6 &	    &	 \\
A 3578	       & 13 57 30.2 & -24 43 00 & I-II  & 3 & 11152 &  5 \\
A 1822	       & 13 58 29.3 & -25 21 59 &	& 6 &	    &	 \\
A 3580	       & 14 02 54.5 & -23 43 25 & I	& 6 &	    &	 \\
A 753S	       & 14 03 38.5 & -33 58 23 & I	& 1 &  4197 &  1 \\
A 1846	       & 14 03 41.5 & -25 21 59 &	& 6 &	    &	 \\
A 3581	       & 14 07 29.7 & -27 01 00 & I	& 3 &  6416 &  1 \\
A 1857	       & 14 08 25.9 & -27 01 15 &	& 6 &	    &	 \\
\tablecomments{Columns: 1. Cluster identification. 2. Right ascension.
3. Declination. 4. Bautz-Morgan type. 5. Abell distance class (ACO).
6. Recession velocity (redshift). 7. Reference for the redshift.}
\tablerefs{1. Taken from the NASA/IPAC Extragalactic Database (NED); 
2. Carrasco et al. (2000: Paper IV); 3. Pierre et al. (1994); 
4. Quintana et al. (1995: Paper I); 5. Quintana et al. (1997); 
6. Stein (1997); 7. Stocke et al. (1991); 8. Vikhlinin et al. (1998).}
\enddata
\end{deluxetable}
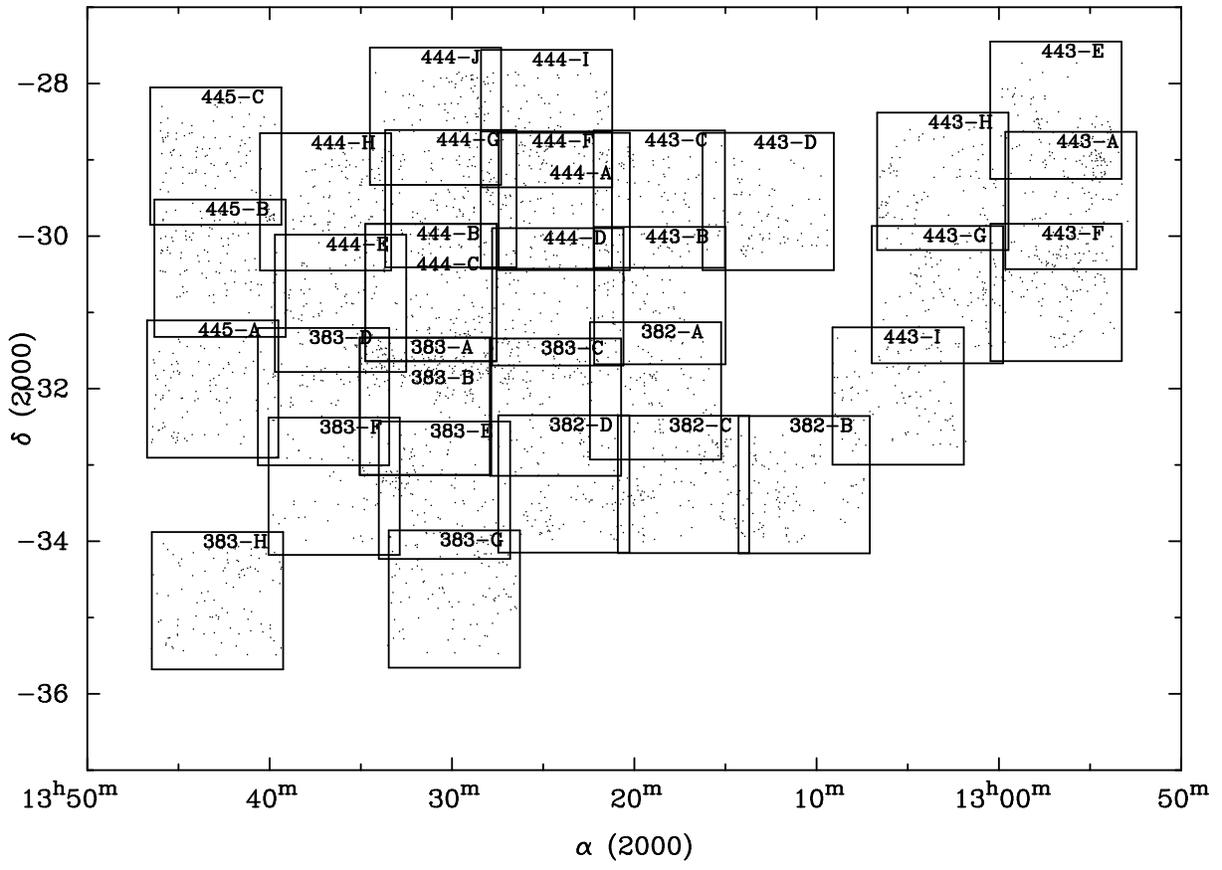
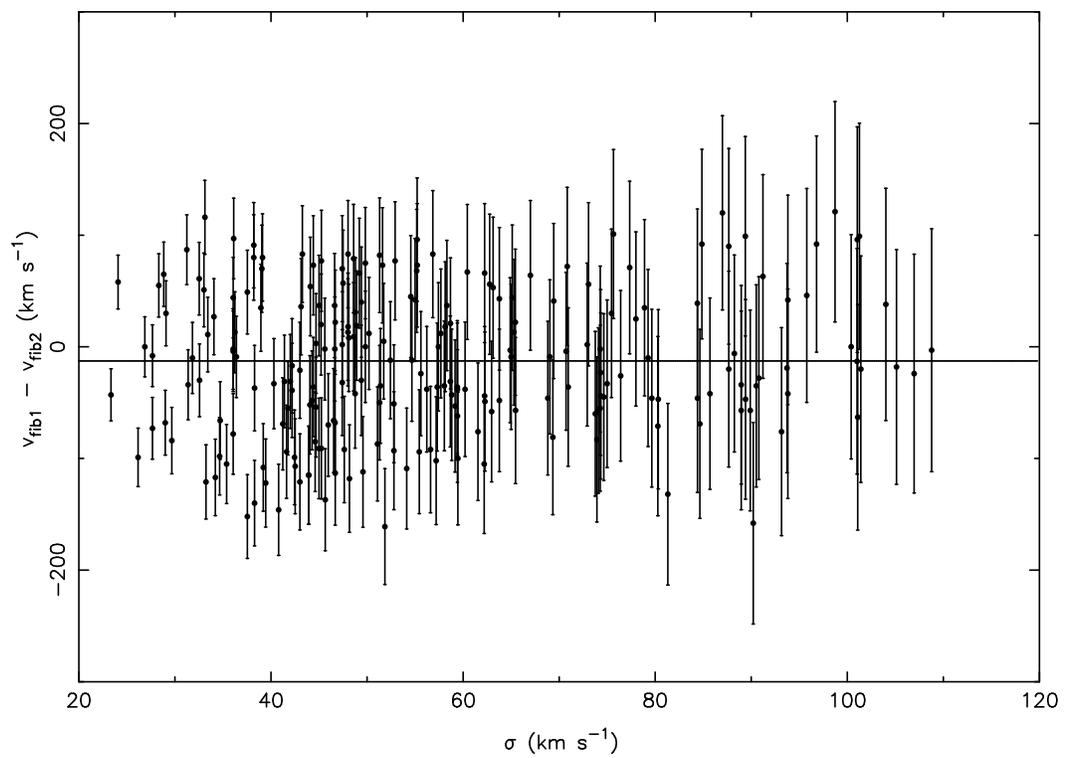
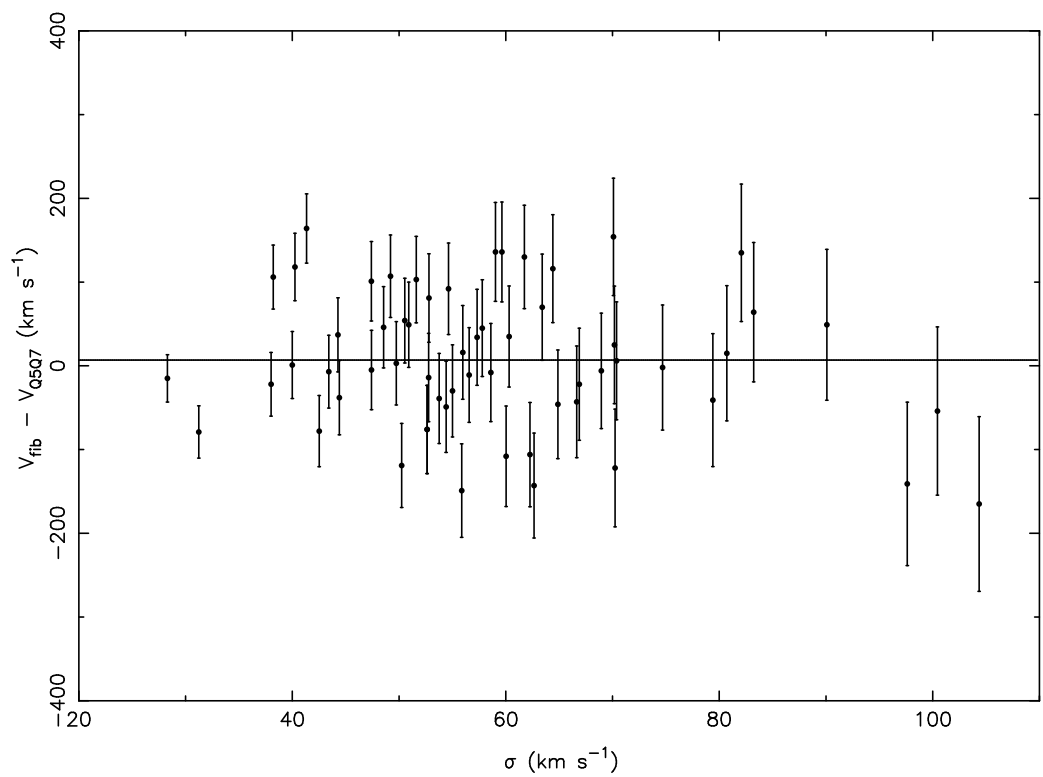
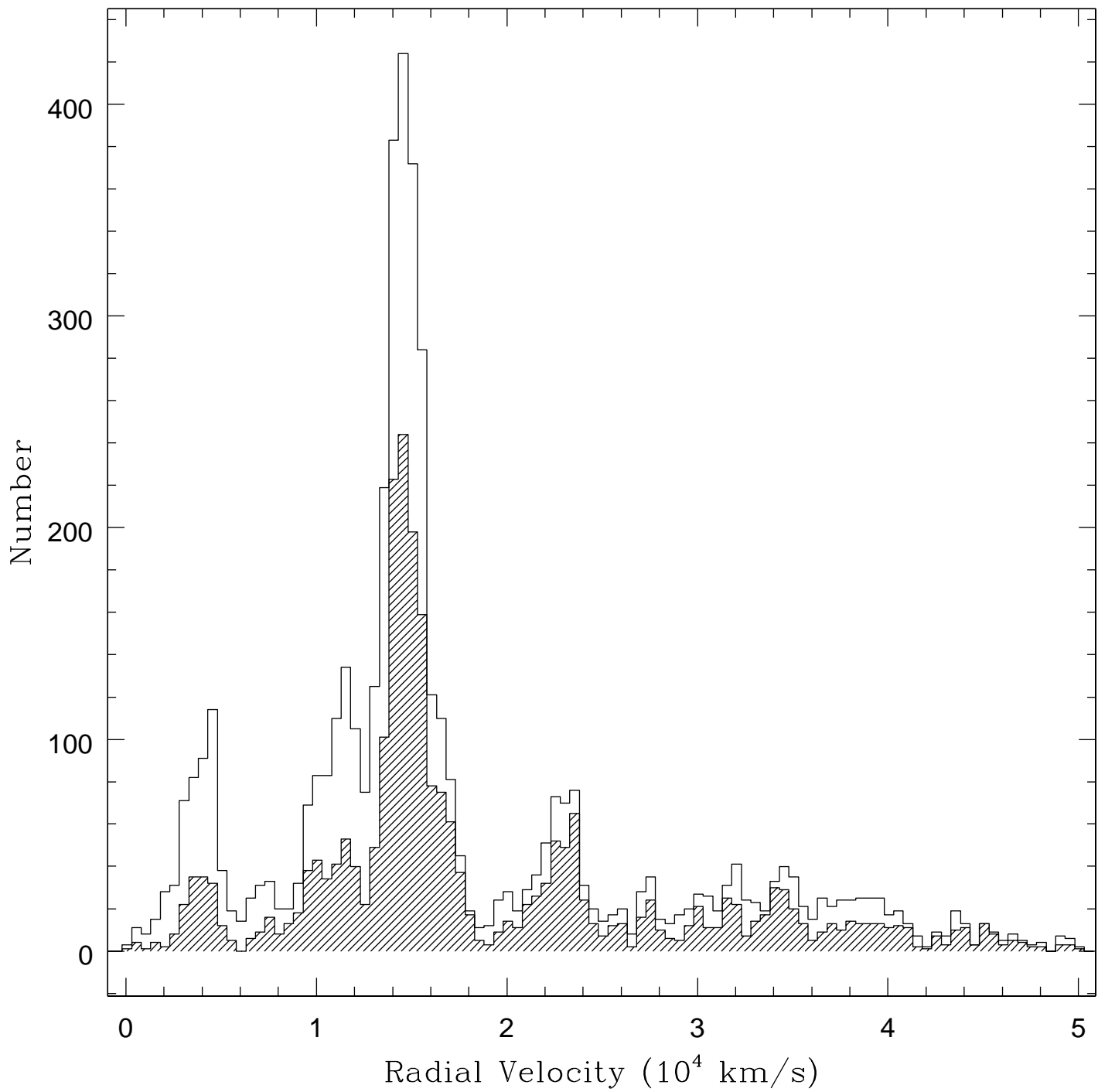
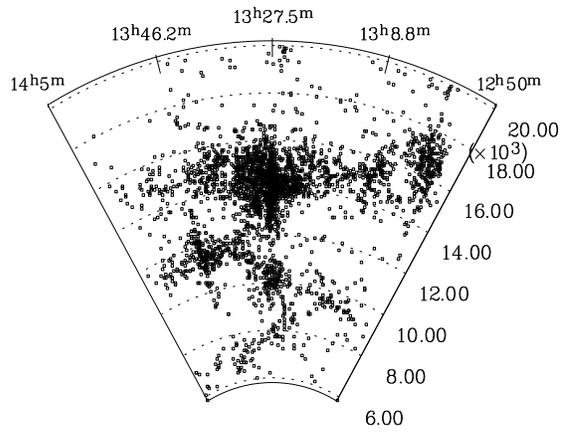
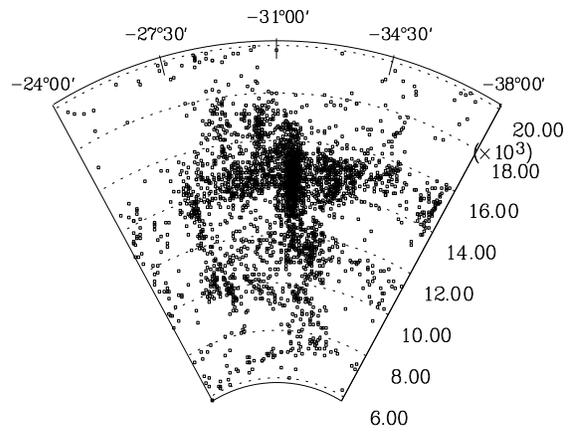
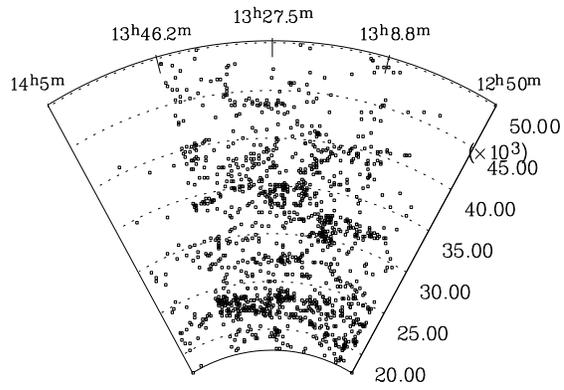
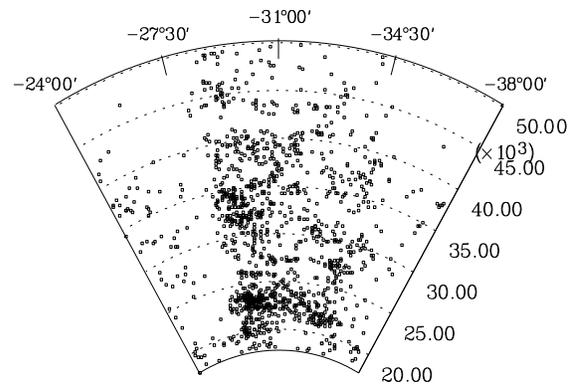
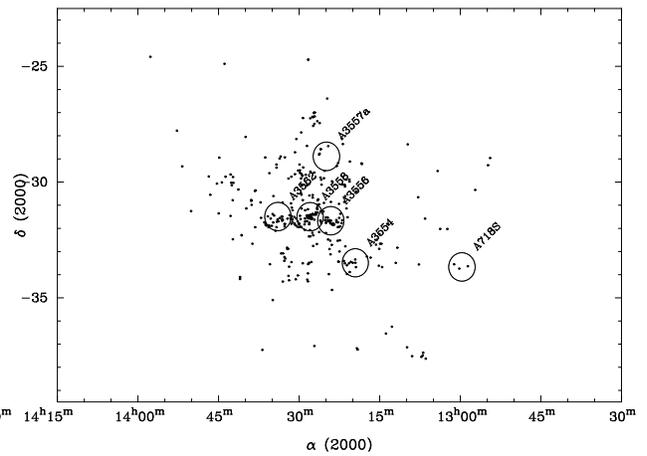
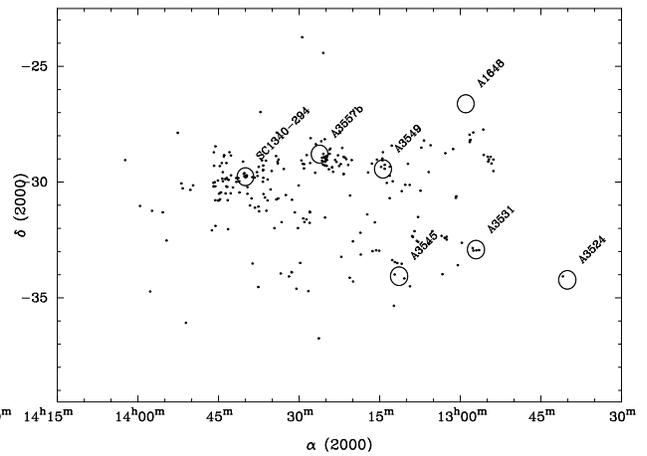
\end{document}